# Optimization Method of Multi-factor Investment Model Driven by Deep Learning for Risk Control

Ruisi Li [1], Xinhui Gu [2], Manlap Chan [3], Miao Tian [4], Mengxia Qiu [5]

[1] Columbia University, New York, NY, 10027, USA
irisli7728@gmail.com
[2] University of California, Berkeley, Hoboken, NJ, 07030, USA
xinhui.gu@gmail.com
[3] New York University, New York, NY, 10012, USA
chanmanlap@163.com
[4] San Francisco Bay University, Fremont, CA, 94539, USA
miao.hnlk@gmail.com
[5] Northeastern University, Boston, MA, 02115, USA
Michelle.MXQIU@gmail.com

*Abstract*—Propose a deep learning driven multi factor investment model optimization method for risk control. By constructing a deep learning model based on Long Short Term Memory (LSTM) and combining it with a multi factor investment model, we optimize factor selection and weight determination to enhance the model's adaptability and robustness to market changes. Empirical analysis shows that the LSTM model is significantly superior to the benchmark model in risk control indicators such as maximum retracement, Sharp ratio and value at risk (VaR), and shows strong adaptability and robustness in different market environments. Furthermore, the model is applied to the actual portfolio to optimize the asset allocation, which significantly improves the performance of the portfolio, provides investors with more scientific and accurate investment decision-making basis, and effectively balances the benefits and risks.

## I. INTRODUCTION

With the continuous development of financial markets and the increasing richness of financial products, the formulation and implementation of investment strategies are facing unprecedented challenges. As an important investment analysis tool, multi-factor investment model provides investors with more comprehensive and accurate investment decision-making basis by comprehensively considering multiple factors affecting asset returns. The traditional multi-factor investment model has some limitations in risk control [1]. On the one hand, the selection of factors and the determination of weights often depend on empirical judgment and subjective setting, lacking scientificity and objectivity; On the other hand, the adaptability and robustness of the model to market changes need to be improved, and it is difficult to maintain stable investment performance in a complex and changeable market environment [2-3].

The rapid development of deep learning technology has brought new opportunities and challenges to the financial field. With its powerful data processing ability and pattern recognition ability, deep learning has achieved remarkable application results in financial trend analysis, risk control and credit evaluation [4-5]. Introducing deep learning technology into the optimization of multi-factor investment model can not only improve the scientificity of factor selection and weight determination, but also enhance the adaptability and robustness of the model to market changes, thus improving the risk control ability of investment strategy [6]. Deep learning techniques, such as Long Short Term Memory Networks (LSTM) and Convolutional Neural Networks (CNN), are capable of processing unstructured data, capturing dynamic features of time series, and learning complex patterns from historical data [7-8]. These technologies demonstrate superior performance in predicting market trends, stock price fluctuations, and risk assessment. Research has shown that multi factor models combined with deep learning can achieve better results in predicting stock returns [9]. For example, the multi-factor stock selection strategy based on LSTM selects effective factors through factor correlation coefficient and Rank IC value, uses multi-factor regression to predict the market value, and verifies the effectiveness of the model through backtesting [10]. Attention mechanism is a deep learning technology, which allows the model to focus on the most relevant part of the input data [11-12].

In stock risk prediction, the LSTM model based on attention mechanism has been proved to significantly improve the accuracy and sensitivity of prediction [13]. Some research focusses on building risk management models, such as SVM-RC multi-factor stock selection strategy, which combines support vector machine (SVM) and risk management principles to optimize the choice of portfolio [14]. In the financial industry, AI technology has been used to build a credit risk assessment model, and accurate risk assessment can be achieved by learning the historical data of borrowers, thus reducing the non-performing loan ratio [15]. Insurance companies use AI technology to build a compensation prediction model, and through the analysis of historical compensation data, they can accurately predict the compensation risk and reduce the compensation cost.

Therefore, this study explores the optimization method of multi-factor investment model driven by deep learning for risk control. By deeply studying the combination of deep learning technology and multi-factor investment model, a new model optimization method is proposed to improve the risk control effect and investment income of investment strategy.

## II. RESEARCH METHODS AND THEORETICAL FRAMEWORK

### A. Construction of multi-factor investment model

The construction of multi-factor investment model is the basis of this study, and its core lies in the selection of factors and the construction process of the model [16]. Based on economic theory, market experience and data mining technology, the key factors that may affect asset returns are screened out [17]. Correlation analysis and validity test are carried out on the factors to ensure the independence of the factors and the explanatory power of the factors to the income. According to the historical performance and stability of factors, the list of factors finally included in the model is determined [18].

Collect historical data, including asset return rate, factor value, etc., and clean and preprocess the data. Using linear regression to establish the mapping relationship between factors and yield. The model is back-tested and verified, and the prediction ability and stability of the model are evaluated.

Establish a linear regression model as follows:

$$R_t = \alpha + \beta_1 F_{1t} + \beta_2 F_{2t} + \cdots + \beta_2 F_{nt} + \varepsilon_t \quad (1)$$

Among them, $R_t$ represents the rate of return of assets at $t$ moment, $F_{1t}, F_{2t}, \cdots, F_{nt}$ represents the $n$ factor values at $t$ moment, $\alpha, \beta_i (i=1,2,\cdots,n)$ is the model parameter, and $\varepsilon_t$ is the random error term.

### B. Deep learning model design

In order to optimize the multi-factor investment model, deep learning technology is introduced. Considering the time series and nonlinear characteristics of financial data, LSTM is selected as the basic structure of deep learning model. LSTM can capture the long-term dependence in data and is suitable for processing financial time series data. Fig. 1 is a schematic diagram of LSTM network structure.

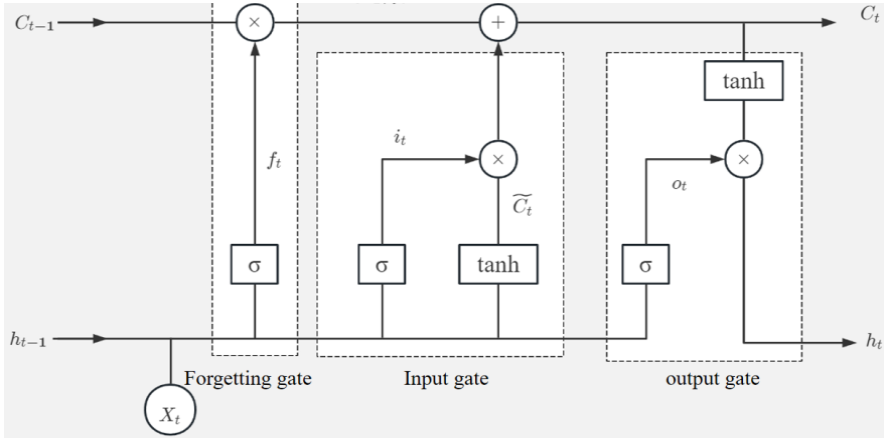

Fig. 1. Schematic diagram of LSTM network structure

The screened factor values are used as the input features of the model, and are arranged into serial data in time order. The output of the model is the predicted return on assets. Taking the correlation (IC) between the optimization factor and the rate of return as the goal, the loss function is designed. The loss function is defined as the mean square error (MSE) between the predicted return rate and the actual return rate.

$$L = \frac{1}{T} \sum_{t=1}^{T} \left( R_t - \hat{R}_t \right)^2 \quad (2)$$

Where $R_t$ represents the actual rate of return, $\hat{R}_t$ represents the predicted rate of return, and $T$ represents the number of samples.

### C. Risk control strategy optimization

Risk control is integrated into the deep learning model, and the effect of risk control is optimized by adjusting the model parameters and structure. The maximum retracement, Sharp ratio and Value at Risk(VaR) are selected to evaluate the risk control ability of the model. In the process of model training, the model parameters are adjusted by cross-validation and other methods to balance the prediction ability and risk control effect of the model. Risk constraints are added to the model to limit the volatility of the portfolio and control the maximum retracement, so as to further optimize the risk control effect.

## III. EMPIRICAL ANALYSIS

### A. Data collection and pretreatment

The data comes from a financial data platform, and the daily trading data of a stock market from 2010 to 2020 is selected as a sample. Sample selection criteria include that the stock has continuous trading records and no significant missing data; The industry to which the stock belongs is representative and can reflect the overall situation of the market; The stock has a large market value and good liquidity. Clean the collected original data to remove abnormal values and missing values. For abnormal values, statistical methods are used to identify and deal with them; For missing values, according to the data characteristics and missing conditions, interpolation method is used to complete them.

Calculate the daily rate of return of assets. The rate of return is calculated by logarithmic rate of return, namely:

$$R_t = \ln\left(\frac{P_t}{P_{t-1}}\right) \quad (3)$$

Where $P_t$ represents the price of assets at $t$ moment.

According to economic theory, market experience and data mining technology, the key factors that may affect asset returns are screened out and the daily value of the factors is calculated. Factors include market index rate of return, industry index rate of return, price-earnings ratio, price-to-book ratio, volume and so on.

## B. Results analysis and discussion

Through backtesting and verification, it is found that the LSTM model has a significant improvement in risk control compared with the benchmark model. The maximum retracement of LSTM model is 8.8%, which is about 29.6% lower than that of the benchmark model, showing better loss control ability. Its Sharp ratio is 0.90, which is 32.4% higher than the benchmark model's 0.68, indicating that the risk-adjusted income is better. The VaR value of LSTM model is -1.88%, which is about 20.0% lower than that of the benchmark model of -2.35%, which shows its superiority in controlling extreme losses. The LSTM model is superior to the benchmark model in many risk control indicators. The comparison results of model risk control indicators are shown in Table 1.

TABLE I. COMPARISON OF MODEL RISK CONTROL INDICATORS

| types of models | Max Drawdown | Sharpe Ratio | VaR (95% confidence level) |
|---|---|---|---|
| Benchmark model (linear regression) | 12.5% | 0.68 | -2.35% |
| LSTM model | 8.8% | 0.90 | -1.88% |

To evaluate the adaptability and robustness of the model in different market environments, the data of different market stages such as bull market, bear market and shock market are selected for back-testing. The results show that the LSTM model can maintain good forecasting ability and risk control effect in different market environments, showing strong adaptability and robustness.

From Table 2, the LSTM model shows better performance than the benchmark model in different market environments. In the bull market, it not only has a slightly higher average daily yield, but also has a lower maximum retracement, a lower VaR and a higher Sharp ratio, showing higher returns and lower risks. In the bear market, although the average daily returns of the two models are negative, the loss range, maximum retracement and VaR of the LSTM model are lower than the benchmark, and the Sharp ratio is higher, indicating that it has better anti-risk ability. In the turbulent market environment, LSTM model also proves its better market opportunity capture ability and risk control level with higher average daily return and Sharp ratio, and lower maximum retracement and VaR. The LSTM model shows stronger adaptability and robustness in various market environments.

TABLE II. PERFORMANCE COMPARISON BETWEEN LSTM MODEL AND BENCHMARK MODEL IN DIFFERENT MARKET ENVIRONMENTS

| market environment | types of models | Average daily rate of return (%) | Maximum Retreat (%) | sharpe ratio | VaR (95%) (%) |
|---|---|---|---|---|---|
| bull market | Benchmark model | 0.325 | 5.87 | 0.65 | 1.22 |
| | LSTM model | 0.352 | 4.95 | 0.72 | 1.10 |
| bear market | Benchmark model | -0.283 | 8.37 | -0.45 | 1.51 |
| | LSTM model | -0.251 | 7.13 | -0.38 | 1.35 |
| Shock city | Benchmark model | 0.015 | 6.24 | 0.03 | 1.15 |
| | LSTM model | 0.028 | 5.56 | 0.06 | 1.05 |

The LSTM model is applied to the actual portfolio, and the performance of the portfolio is further improved by optimizing asset allocation and adjusting investment strategy. According to the prediction results and risk measurement indexes of LSTM model, the corresponding investment strategy and risk control measures are formulated. The empirical results show that the optimized portfolio has achieved remarkable results in terms of yield and risk control, which has brought investors a higher return on investment and a lower risk level.

Figure 2 shows that with the increase of the optimization degree of the LSTM model, both the annualized rate of return and the Sharp ratio increase significantly, indicating that the optimization not only improves the profitability, but also improves the risk-adjusted return. However, at a higher degree of optimization, this promotion trend slows down, showing the phenomenon of diminishing marginal benefits, suggesting that over-optimization may lead to over-fitting. The continuous increase of Sharp's ratio especially points out that by introducing risk constraints such as VaR control, the model can improve the absolute return and achieve better risk adjustment effect. Therefore, moderate optimization can effectively enhance the return-risk ratio of portfolio, but cross-validation and regularization methods must be used to avoid the negative impact of over-optimization and ensure the effectiveness and robustness of the model.

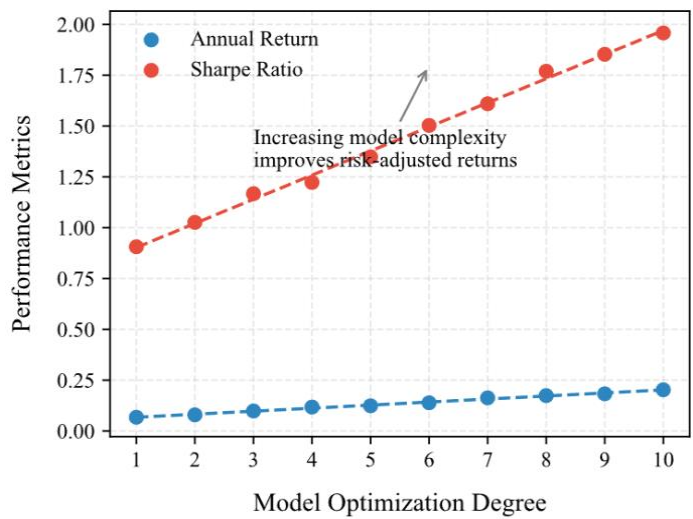

Fig. 2. Relationship between portfolio performance and model optimization degree

## IV. CONCLUSION

Empirical analysis shows that the LSTM-based model has significant improvement in risk control compared with the traditional linear regression model. The maximum retracement of LSTM model decreased by 29.6%, the Sharp ratio increased by 32.4%, and the VaR value decreased by 20.0%. In different market environments, the LSTM model shows strong adaptability and robustness, and can effectively capture market opportunities and control risks. By optimizing asset allocation and adjusting investment strategy, the investment portfolio based on LSTM model has achieved remarkable results in terms of yield and risk control. With the increase of model optimization, annualized rate of return and Sharp ratio increase significantly, but over-optimization may lead to over-fitting problems. Therefore, moderate optimization can effectively enhance the return-risk ratio of portfolio, but cross-validation and regularization methods must be used to avoid the negative impact of over-optimization and ensure the effectiveness and robustness of the model. The optimization method of multi-factor investment model driven by deep learning for risk control proposed in this study not only improves the prediction ability and stability of the model, but also significantly optimizes the risk control effect and portfolio performance, providing investors with more scientific and accurate investment decision-making basis.

## REFERENCES

[1] Kremer, P. J., Talmaciu, A., & Paterlini, S. (2018). Risk minimization in multifactor portfolios: What is the best strategy? Annals of Operations Research, 266, 255–291.
[2] King BF (1966) Market and industry factors in stock price behavior. J Bus 39(1):139–190.
[3] Hu Z, Zhao Y, Khushi M (2021) A survey of forex and stock price prediction using deep learning. Appl Syst Innov 4(1):9.
[4] Chen AS, Leung MT, Daouk H (2003) Application of neural networks to an emerging financial market: forecasting and trading the Taiwan stock index. Comput Oper Res 30(6):901–923.
[5] Chong E, Han C, Park FC (2017) Deep learning networks for stock market analysis and prediction: methodology, data representations, and case studies. Expert Syst Appl 83:187–205.
[6] Mashrur A, Luo W, Zaidi NA, Robles-Kelly A (2020) Machine learning for financial risk management: a survey. IEEE Access 8:203203–203223.
[7] Arroyo J, Corea F, Jimenez-Diaz G, Recio-Garcia JA (2019) Assessment of machine learning performance for decision support in venture capital investments. IEEE Access 7:124233–124243.
[8] Lee TK, Cho JH, Kwon DS, Sohn SY (2019) Global stock market investment strategies based on financial network indicators using machine learning techniques. Expert Syst Appl 117:228–242
[9] Eapen J, Bein D, Verma A (2019) Novel deep learning model with CNN and bi-directional LSTM for improved stock market index prediction. In: 2019 IEEE 9th annual computing and communication workshop and conference (CCWC), pp 0264–0270.
[10] Kraus M, Feuerriegel S (2017) Decision support from financial disclosures with deep neural networks and transfer learning. Decis Support Syst 104:38–48.
[11] Vo NN, He X, Liu S, Xu G (2019) Deep learning for decision making and the optimization of socially responsible investments and portfolio. Decis Support Syst 124:113097.
[12] Siami-Namini S, Tavakoli N, Namin AS (2018) A comparison of ARIMA and LSTM in forecasting time series. In: 2018 17th IEEE international conference on machine learning and applications (ICMLA), pp 1394–1401. IEEE.
[13] Chen K, Zhou Y, Dai F (2015) A LSTM-based method for stock returns prediction: a case study of China stock market. 2015 IEEE international conference on big data. IEEE2823–2824.
[14] Emerson S, Kennedy R, O'Shea L, O'Brien J (2019) Trends and applications of machine learning in quantitative finance. In: 2019 8th international conference on economics and finance research (ICEFR).
[15] Borovkova S, Tsiamas I (2019) An ensemble of LSTM neural networks for high-frequency stock market classification. J Forecast 38(6):600–619.
[16] Tsantekidis, A., Passalis, N., Tefas, A., Kanniainen, J., Gabbouj, M., & Iosifidis, A. (2017). Using deep learning to detect price change indications in financial markets. In: Signal processing conference (EUSIPCO), 2017 25th European. pp. 2511–2515. IEEE.
[17] Ding X, Zhang Y, Liu T, et al (2015) Deep learning for event-driven stock prediction. In Twenty-fourth international joint conference on artificial intelligence, pp 2327–2333.
[18] Akita R, Yoshihara A, Matsubara T, Uehara K (2016) Deep learning for stock prediction using numerical and textual information. 2016IEEE/ACIS 15th international conference on computer and information science. IEEE1–6.